# Giant increase in the metal-enhanced fluorescence of organic molecules in nanoporous alumina templates and large molecule-specific red/blue shift of the fluorescence peak

S. Sarkar[1], B. Kanchibotla[2], J. D. Nelson[3], J. D. Edwards[3], J. Anderson[3], G. C. Tepper[1] and S. Bandyopadhyay[2]

[1]Department of Mechanical and Nuclear Engineering, Virginia Commonwealth University, Richmond, Virginia 23284, USA

[2]Department of Electrical and Computer Engineering, Virginia Commonwealth University, Richmond, Virginia 23284, USA

[3]US Army Engineer Research and Development Center, Alexandria, Virginia 22315, USA

## ABSTRACT

The fluorescence of organic fluorophore molecules is enhanced when they are placed in contact with certain metals (Al, Ag, Cu, Au, etc.) whose surface plasmon waves couple into the radiative modes of the molecules and increase the radiative efficiency. Here, we report a hitherto unknown size dependence of this metal enhanced fluorescence (MEF) effect in the nanoscale. When the molecules are deposited in nanoporous anodic alumina films with exposed aluminum at the bottom of the pores, they form organic nanowires standing on aluminum nanoparticles whose plasmon waves have much *larger* amplitudes. This increases the MEF strongly, resulting in several *orders of magnitude* increase in the fluorescence intensity of the organic fluorophores. The increase in intensity shows an inverse *super-linear* dependence on nanowire diameter because the nanowires also act as plasmonic "waveguides" that concentrate the plasmons and increase the coupling of the plasmons with the radiative modes of the molecules. Furthermore, if the nanoporous template housing the nanowires has built-in electric fields due to space charges, a strong *molecule-specific* red- or blue-shift is induced in the fluorescence peak owing to a renormalization of the dipole moment of the molecule. This can be exploited to detect minute amounts of target molecules in a mixture using their optical signature (fluorescence) despite the presence of confounding background signals. It can result in a unique new technology for bio- and chemical-sensing.

Detection of trace amounts of biological and chemical agents in the environment is a challenging task since any signal from the target is usually masked by background signals. This is particularly true in the case of optical detection where the fluorescence of the target has to be sensed and identified unambiguously amidst confounding signals due to other agents. Unfortunately, the fluorescence spectrum of the target species may (1) overlap with the broad fluorescence spectrum of the background, and (2) may be weaker than the background fluorescence, so the target remains undetectable. A possible remedy is to induce a large red- or blue-shift in the fluorescence spectrum of the target molecules selectively while leaving the fluorescence spectrum of the background unaffected. This may allow separating out the fluorescence signal of the target from the background by performing frequency-selective detection. Furthermore, if we can also increase the fluorescence efficiency of the target species selectively, then we can make the target signal stronger and more easily detectable. In this Letter, we report a series of experiments that have achieved *both* goals, i.e. inducing a molecule-specific red- or blue-shift in the fluorescence peak of target molecules to improve the "selectivity" of detection, and simultaneously increasing the fluorescence efficiency of the target to improve the "detectivity".

We prepared a set of test samples by electro-spraying target molecules inside nanometer-sized pores of ~ 1 μm thick anodic alumina films which were produced by anodizing 99.994% pure aluminum foils (0.1 mm thick) in 15% sulfuric or 0.3M oxalic acid using different voltages[1]. Prior to anodization, the foils were electropolished in a solution of perchloric acid, butyl cellusolve, ethanol and distilled water to reduce the surface roughness to about 3 nm[1, 2]. The process of anodization involved placing the aluminum foil in contact with the acid and then passing a dc current through the acid using the foil as the anode and a platinum mesh as the cathode. This process created a nanoporous film of alumina on the surface of the aluminum foils.

The pore diameters in the nanoporous films depend on the acid used and the dc voltage employed during anodization. Anodization in sulfuric acid produces pores of diameter 10 nm, while anodization in oxalic acid produces pores of diameter 20 nm if the anodizing voltage is 25 V, and 50 nm if the anodizing voltage is 40 V[3]. At the bottom of the pores, there is a "barrier layer" of alumina in contact with the aluminum substrate. This layer is first removed using 'reverse polarity etching'[4] to expose the aluminum at the bottom of the pores. Next, the organic molecules are dissolved in a liquid solvent and electro-spray ionization[5-8] is used to deposit them *selectively* within the pores. In the electro-spray process, charged droplets are generated at the tip of a metal needle (or pipette with a wire immersed in the liquid) and are subsequently delivered to a grounded target (e.g. the alumina film). The droplets are derived by charging a liquid typically to 5-20 kV vs. the target, which leads to charge injection into the liquid from the electrode. The charged liquid is attracted to the ground electrode of opposite polarity, forming a so-called Taylor cone at the needle tip. Droplets are formed when electrostatic forces between the charged liquid and the ground exceed the liquid's surface tension. If the liquid is relatively volatile, the solvent liquid will evaporate and a dry stream of solute ions will be deposited onto the target. In the case of the nanoporous alumina target, the insulating alumina regions will quickly charge up until the charge density on the alumina surface produces an electric field equal and opposite to the externally applied electric field. The surface charge on the alumina template will force the electric field lines to penetrate into the pores and terminate on the back metal contact. As a result, the electro-sprayed species is selectively deposited *only within the pores* and not outside it (see Supporting Information).

The fluorescence spectra of the samples obtained at room temperature showed the following features:

- The fluorescence efficiency of molecules increases many fold when they are confined within pores with exposed metal at the bottom. The increase is monotonic with decreasing pore diameter and has a strong super-linear dependence on inverse of pore diameter.
- There is a giant red- or blue-shift in the fluorescence peak of molecules within the pores. Both the magnitude and the sign of the shift is molecule-specific, i.e. different molecules experience different relative shifts $\Delta\lambda / \lambda$ ($\lambda$ is the wavelength corresponding to the fluorescence peak). This molecular specificity allows us to discriminate between different molecules and provides the means for molecular recognition.

The former feature is caused by the well-known phenomenon of "metal-enhanced-fluorescence" (MEF)[9-25] where surface plasmon fields, originating from coherent electron density fluctuation at the interface of the fluorophore and exposed aluminum at the pore bottom, couple into the organic fluorophore and increase the radiative recombination rate of photogenerated electron-hole pairs. This, in turn, increases the quantum yield of photons from the recombination process and therefore increases the fluorescence efficiency of the fluorophore in the pores. The amplitude of the surface plasmon field will be strongest in the narrowest pores since the fluorophores in these pores have the smallest interface with the metal at the bottom. In the smallest diameter pores, a fixed fluctuation $\Delta N$ in the electron population at the interface results in the largest surface density fluctuation $\Delta N / A$ where $A$ is the area of the interface. The smallest pores therefore experience the strongest surface plasmon field, so the increase in the fluorescence efficiency should be largest in the smallest pores. It is also quite possible that the surface plasmon resonance bands in the smallest pores, with the smallest aluminum nanoparticles at the bottom, have the highest overlap with the emission or excitation spectra of the fluorophores, resulting in the strongest coupling in the smallest pores and hence the strongest enhancement of luminescence there[25].

The physics of 'metal-enhanced-fluorescence' has been discussed extensively in ref. [11]. The plasmon wave has a far-field component and a near-field component that decays rapidly with distance (evanescent component). The far-field component can propagate into the pores only if correct wavevector matching occurs[11], which is why it is more likely that the wave that enters the pores is mostly evanescent. The larger the amplitude of the evanescent wave ($E_0$) that enters the pores, the larger is the plasmon density at a given distance from the metal-fluorphore interface since the evanescent wave decays with distance $z$ as $E_0 e^{-\kappa z}$ (where $\kappa$ is the imaginary part of the wavevector) and the plasmon density varies as $E_0^2 e^{-2\kappa z}$ . A plasmon stimulates an excited dipole in the fluorophore to relax to a lower energy state by emitting a photon. This is similar to stimulated emission of light in lasers. If the amplitude of the plasmon wave that enters the pores is larger, a larger fraction of the fluorophore molecules in the pores experience the plasmon stimulation from the evanescent wave and hence the radiation output is larger. That is why the pores with the smallest diameter should radiate the most and exhibit the highest fluorescence efficiency.

In Fig. 1, we show the fluorescence spectra of Rhodamine 123 molecules electro-sprayed inside pores as a function of pore diameter. The pore diameters were 10, 20, 30 and 50 nm. In this measurement, the excitation intensity was kept constant and the amount (or volume) of deposited material, determined by the duration of electro-spraying (30 minutes), was also kept constant. Therefore, the measured intensity of fluorescence is directly proportional to the fluorescence efficiency, which clearly increases with decreasing pore diameter. Note that the increase is not gradual; the fluorescence efficiency increases rapidly once the pore diameter shrinks below 30 nm. The increase is *much stronger* than what would be predicted by the inverse dependence on pore cross sectional area (recall that the amplitude of the surface plasmon field is proportional to $\Delta N / A$ and hence the plasmon density at any given location inside a pore should have been inversely proportional to $A^2$). Therefore, it is not just the usual MEF, but an additional effect is at play. The fact that the increase is much steeper than predicted by the inverse dependence on pore area suggests that it is not just the field amplitude that is increasing in narrower pores, but also the *coupling* of the field with the radiative modes of the molecules is stronger. This will happen if the organic nanowires are acting as plasmonic waveguides that concentrate the plasmons and hence increase the coupling. Nanowires with narrower diameter will sustain fewer coupled light-plasmon (plasmon-polariton) modes[26] while larger diameter ones will sustain more modes. Since multi-mode propagation is more lossy than monomode propagation, the narrower pores will be able to propagate the confined polaritons over longer distances[27] and hence will be more effective in enhancing fluorescence. To our knowledge, this waveguiding effect on MEF has not been reported before.

It might have been tempting to explain the rapid increase of fluorescence with decreasing nanowire diameter (the data in Fig. 1) by invoking an alternate explanation, namely quantum confinement of photogenerated electron-hole pairs within the pores. That would have been appropriate for *inorganic* semiconductors where the wavefunctions of photogenerated electrons and holes are delocalized and spread out over the width of the nanowire. Consequently, the pore walls would have confined the electron and hole wavefunctions and increased the overlap between them, resulting in an increase in the fluorescence efficiency that would become stronger with decreasing nanowire diameter. This picture, however, is inappropriate for *organic molecules*, where the wavefunctions are strongly localized over individual atoms or molecules that are much smaller than the pore diameter. Therefore, the pore walls do not cause any additional confinement and quantum confinement cannot be the origin of the fluorescence increase. This leaves MEF and the associated waveguiding effect as the likely cause of what we observe.

In order to gain further evidence in support of the MEF effect as the source of the increase in fluorescence efficiency, we electro-sprayed the same molecules on large-area aluminum films so that the interface between the fluorophore and metal became macroscopically large. There was no visible increase in the fluorescence efficiency in this case, showing that the interface area has to be mesoscopic (nanometer-sized) for the MEF effect to be discernible. This is consistent with surface plasmon resonance. The surface density fluctuation $\Delta N / A$ is negligible when $A$ is macroscopically large. Therefore, the surface plasmon effect has negligible influence when the molecules are deposited on large area metal films, but has a strong influence when the molecules are deposited on nanometer sized metal interfaces.

Finally, in order to confirm the MEF effect as the dominant, if not the sole, cause of intensity increase, we evaporated ~ 1 μm thick aluminum layer on a conducting $n^+$-silicon wafer and then anodized the layer *completely* to produce a porous alumina film. All of the aluminum is converted to alumina in this process, so that in the end, the porous film is supported on the silicon substrate with *no aluminum present*. The barrier layer was removed and Coumarin was electrosprayed within the pores. This time, there was no significant increase in the fluorescence intensity because of the absence of the aluminum interface which is responsible for surface plasmons. Only “free electron” metals (Au, Ag, Cu and Al) that have a single electron in the valence band cause the surface plasmon effect; silicon does not. Therefore, the absence of the intensity increase when aluminum is replaced by silicon confirms that the MEF effect is the source of the intensity increase.

In Fig. 2(a), we compare the fluorescence spectrum of bulk Rhodamine samples with that obtained from Rhodamine electro-sprayed in 20-nm diameter pores. There is a clear red-shift of 30 nm in the peak wavelength when the Rhodamine is confined within pores. In Fig. 2(b), we show the red-shift in the spectrum of a different molecule - Courmarin 314 - electro-sprayed within 20-nm pores. The relative red-shift $\Delta\lambda / \lambda$ ( $\lambda$ is the peak fluorescence wavelength of a bulk sample) is much larger in Coumarin (16%) than in Rhodamine (5.5%). Therefore, the magnitude of the red-shift is *molecule-specific*.

The shift in the peak wavelength occurs because of the built-in electric fields that exist in the alumina host. During the process of electro-spraying, fixed charges are implanted in the surface of the alumina films, which result in a strong dc electric field along the length of those pores that have exposed metal at the bottom (the metal is grounded during the electro-spraying process so that there can be no charge left on the metal). This field alters the dipole moment of the molecule[28] that is contained within the pore. Charged surface states at the fluorophore/alumina interface can also alter the dipole moment. That, in turn, changes the energy gap between the highest occupied molecular orbital (HOMO) and lowest unoccupied molecular orbital (LUMO) levels, causing a blue- or red-shift in the fluorescence peak depending on whether the energy gap expands or shrinks. Obviously the sign (red or blue) and the magnitude of the shift will be molecule-specific since the energy gap will increase in some molecules and decrease in others (depending on their dipole moments). That is exactly what we see. This molecular-specificity can be exploited for bio- and chemical sensing.

To confirm that fixed charges and the associated Stark effect are indeed responsible for the red-shift, we electro-sprayed Coumarin molecules both within the pores and outside the porous film onto bulk aluminum where no electric field can be present because the metal is an excellent conductor that will electrically short out any electric field. Fig. 3 shows an image acquired with a digital camera of the fluorescence. The central circular region is where the aluminum was anodized and contains the porous film (with 20 nm diameter pores). The fluorescence from the molecules confined within the pores is red-shifted to greenish-yellow, while the fluorescence from the molecules deposited on bulk metal (outside the circular region) remains in the blue-violet region corresponding to the natural fluorescence of Coumarin.

The molecule specific red- and blue-shifts can also be used to resolve different constituents in a mixture, even if they have overlapping emission spectra (and are therefore optically indistinguishable) in a neutral environment. In Fig. 4(a) and 4(b), we show the emission spectra of bulk tryptophan and tyrosin along with the spectra obtained when the molecules are electro-sprayed within 50-nm diameter pores. The spectra of bulk molecules overlap in the frequency domain, making trypotophan and tyrosin optically indistinguishable. However, when a mixture of these two biomolecules (in equal molar parts) are electro-sprayed within 50 nm pores of an alumina film, we induce a 40 nm red-shift in the peak wavelength of tryptophan and 120 nm red-shift in the peak wavelength of tyrosin. As a result, we see a double peak structure in the fluorescence spectra of 50-nm samples (Fig. 4(c)) where the high energy peak is easily identifiable with tryptophan and the low energy peak with tyrosin. In bulk form, the mixture does not exhibit a double peak structure as seen in Fig. 4(c), but when it is electro-sprayed into 50 nm pores, the double peak structure immediately shows up. Thus, we have demonstrated a powerful technique that allows us to resolve these otherwise indistinguishable compounds using fluorescence spectra. This can lead to a new technology for bio- and chemical-sensing.

In Fig. 5, we show the normalized emission spectra of dipicolinic acid (DPA) molecules electrodeposited in 10, 20 and 50 nm pores. The 10- and 50-nm pore-diameter samples show no significant shift in the peak frequency (from that of bulk DPA molecules), but the 20 nm pore-diameter samples show a significant *blue-shift* of 55 nm in the peak wavelength. Here, the electric field expands the HOMO-LUMO gap (whereas in Rhodamine and Coumarin it shrinks it) and therefore induces a blue shift in the fluorescence spectrum, as opposed to a red-shift. This shows that not only the magnitude, but also *the sign* of the shift is molecule-specific. The 10- and 50-nm pore-diameter samples show no significant shift because of two different reasons. In the case of the 10-nm pores, the reverse polarity etching is often very incomplete because the pore size is too small to remove the barrier layer from the bottom easily. The 10-nm pore-diameter samples therefore may not have a built-in electric field in large areas of the film and consequently may not exhibit a significant frequency shift. In the case of 50 nm samples, the metal at the pore bottom is easily exposed, but the *average* electric field within the pores is weakest. This happens because the electric field is strongest near the pores walls (since the charges are implanted in the alumina) and weakest at the center of the pores. Therefore, the 50-nm samples will experience the weakest *average* electric field (averaged over the pore volume) and experience the least red- or blue-shift. This explains why molecules within the 50-nm pores show little or no blue shift, but the molecules within 20-nm pores show a strong blue-shift. DPA is an important constituent in bacillus spores (e.g. anthrax) and the ability to engineer its emission spectrum controllably by spraying it into narrow pores provides an important tool for detection and identification of DPA.

In conclusion, we have demonstrated that confining organic fluorophore molecules within anodic alumina pores that have exposed aluminum nanoparticles at the bottom increases the fluorescence intensity and induces molecule specific giant frequency shifts in the fluorescence peak. The former is due to the metal-enhanced fluorescence effect increased several fold by waveguiding of surface plasmon modes. The latter is due to built-in electric fields in the host template that alters the dipole moment and therefore the HOMO-LUMO gap of the molecules. While the former increases the detectivity of target molecules in optical detection schemes, the latter increases the selectivity of molecule sensors based on fluorescence. The molecule specific frequency shifts in the fluorescence spectrum also allows us to deconvolve the spectra of two different constituents in a mixture that have overlapping spectra. This can be used for molecular recognition, and ultimately bio- or chemical-detection.

## ASSOCIATED CONTENT

Supporting data, figures and plots are included in the supporting material. This material is available free of charge via the Internet at http://pubs.acs.org.

## AUTHOR INFORMATION

Corresponding author: S. Bandyopadhyay, sbandy@vcu.edu

S. Sarkar currently at Eaton Corporation, Southfield, Michigan 48076.

B. Kanchibotla, currently at First Solar, Toledo, Ohio 43604.

## Notes

The authors declare no competing financial interests.

**Figure captions**

**Fig. 1**: Fluorescence intensity of organic molecules confined in nanopores as a function of pore diameter. The Rhodamine 123 molecules were electrosprayed into nanopores in anodic alumina films with exposed aluminum at the bottom and also onto a large area aluminum film ('blank') for 30 minutes, resulting in a ~30 nm film on the blank aluminum film. The excitation intensity was kept fixed while measuring the fluorescence intensity. The intensity (or luminescence efficiency) increases monotonically with decreasing pore diameter because the metal nanoparticle (exposed Al) at the bottom of the pore decreases in size with decreasing pore diameter. This causes the amplitude of the surface plasmon wave originating from the Al nanoparticle to increase with decreasing pore diameter. Insofar as these surface plasmons enhance the fluorescence efficiency, the smallest pores will exhibit the largest increase in the fluorescence. Note that the increase in fluorescence not gradual; the intensity increases rapidly as the pore diameter shrinks below 20 nm. This is caused by the nanowires acting as plasmonic waveguides. The narrower wires support fewer waveguide modes and hence are less 'leaky' or 'lossy' so the coupled light-plasmon polaritons can travel further in narrower wires, resulting in the steep super-linear increase of intensity with decreasing pore diameter. The intensity emitted by 10-nm diameter nanowires is ~350 times stronger than that emitted by 50-nm diameter nanowires and several orders of magnitude stronger than that emitted by molecules electrosprayed on large area Al films.

**Fig. 2:** Emission spectra of organic fluorophores in pores showing giant red shifts**.** Emission spectrum of (a) Rhodamine 123 and (b) Coumarin 314 in pores of diameter 20 nm compared to the emission spectrum of the corresponding bulk specimens. The 20-nm diameter Rhodamine nanowires formed inside the pores show a red shift of 30 nm and the 20-nm Coumarin nanowires show a red shift of 80 nm in the peak emission wavelength because of the renormalization of the dipole moments of these molecules due to strong electric fields present in the porous host matrix. This renormalization shrinks the LUMO-HOMO gap of the molecules and causes the red-shift. In the case of Coumarin, the red-shift is giant (>14%).

**Fig. 3**: Digital photograph of fluorescence of molecules deposited in nanopores (central circular region) and bare aluminum films. (a) The molecules confined in pores fluoresce with enhanced intensity (detectable with the naked eye) and the fluorescence is significantly red-shifted in frequency. Note that there are blue streaks in the central circular region. In these streak areas, the electric field has been neutralized (the pore walls discharged), so that there is no red-shift in the fluorescence here. This figure establishes that the pores play an essential role since without them we do not see significant frequency shifts or intensity increases.

**Fig. 4**: Resolving the spectra of two different molecules whose spectra overlap in a neutral environment. (a) The emission spectra of tryptophan in bulk and in 50 nm diameter pores, (b) the emission spectra of tyrosin in bulk and in 50 nm diameter pores, (c) the emission spectra of equal molar parts of tryptophan and tyrosin showing a single peak structure in bulk but a double peak structure in 50 nm pores, where each peak can be associated with a single constituent, resulting in spectral deconvolution. The deconvolution was made possible by the fact that the spectra of the two species experienced very different amounts of red-shift in the nanoporous template.

**Fig. 5:** Blue shift in molecular spectra**.** The normalized emission spectra of dipicolinic acid (DPA) molecules electrodeposited in 10, 20 and 50 nm pores. The 10- and 50-nm pore- diameter samples show no significant shift in the peak frequency (from that of bulk DPA molecules), but the 20 nm samples, show a significant *blue-shift* of 50 nm, possibly because the electric field is much stronger in the 20-nm pore-diameter samples. This example shows that the electric fields in the porous host can expand the LUMO-HOMO gap in some molecules (blue-shift) and shrink the gap in some other molecules (red-shift).

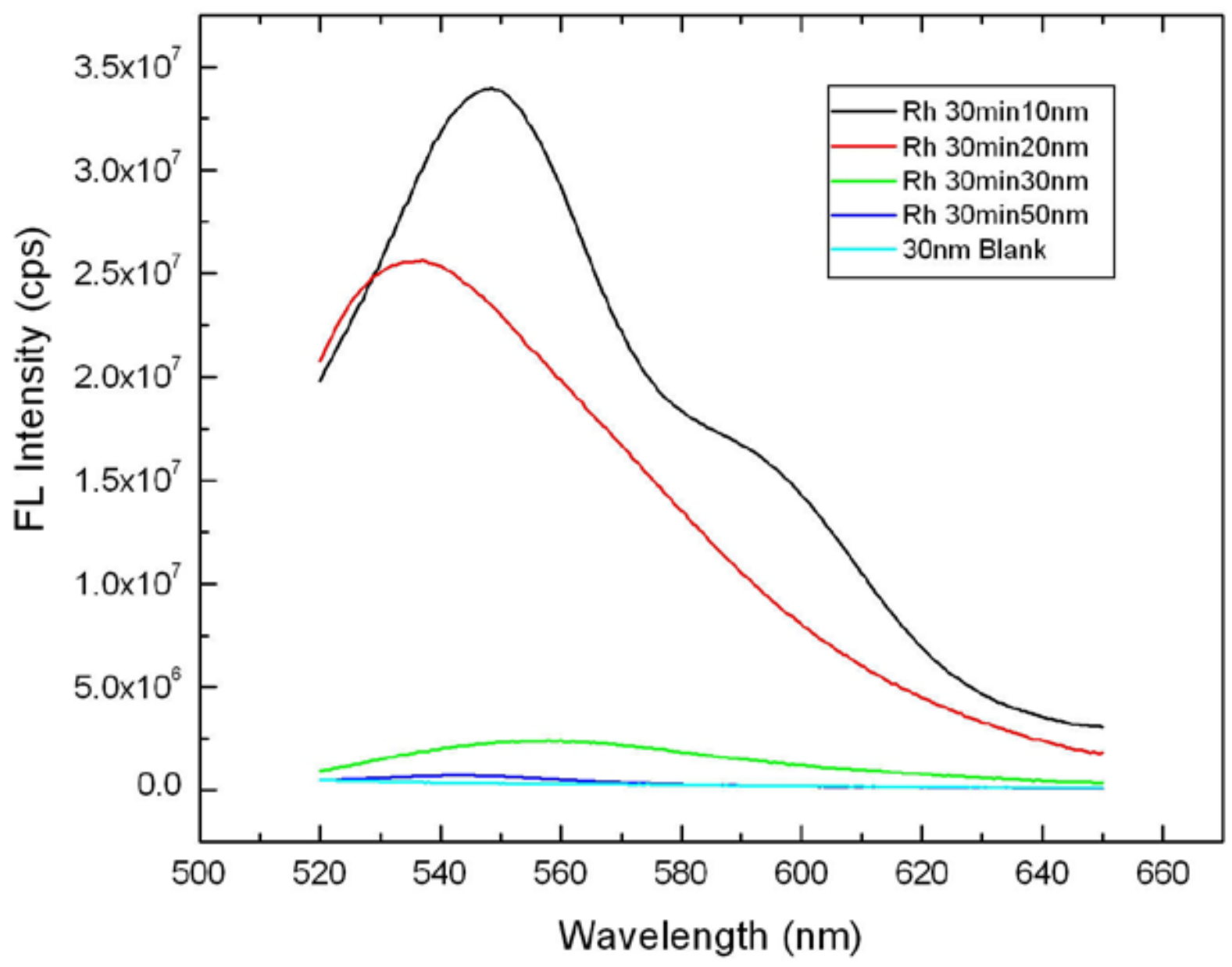

Fig. 1

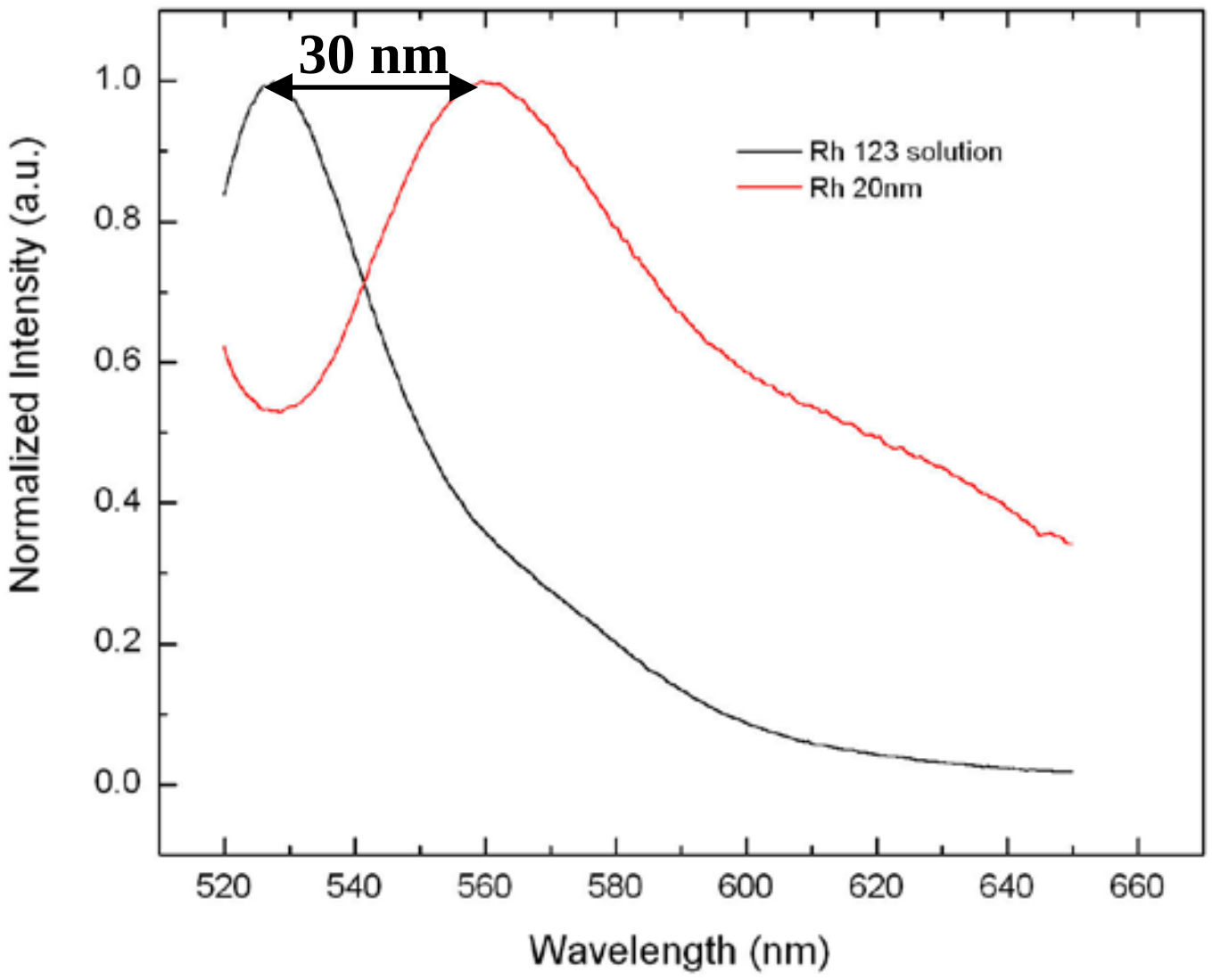

**(a)**

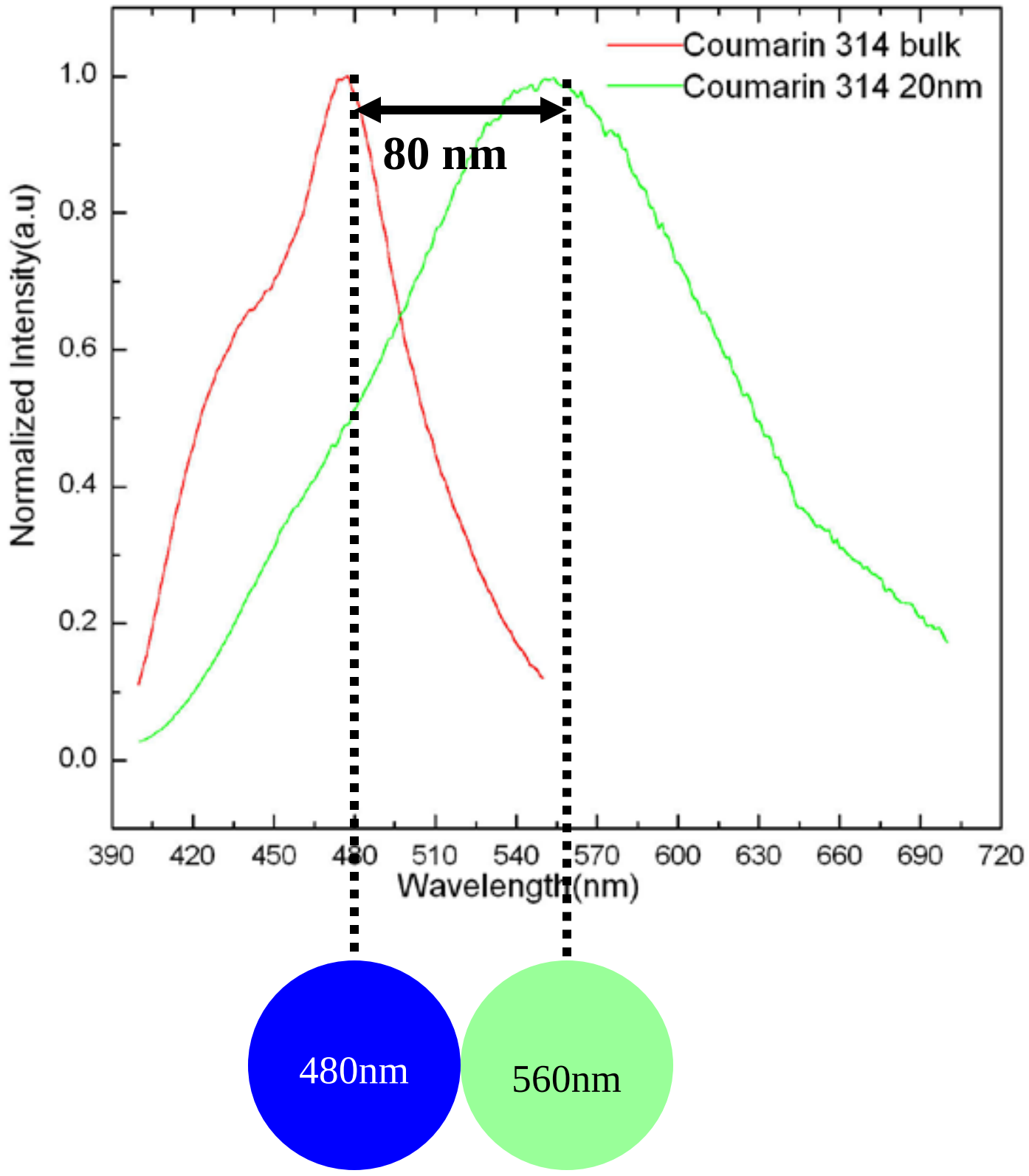

**(b)**

Fig. 2

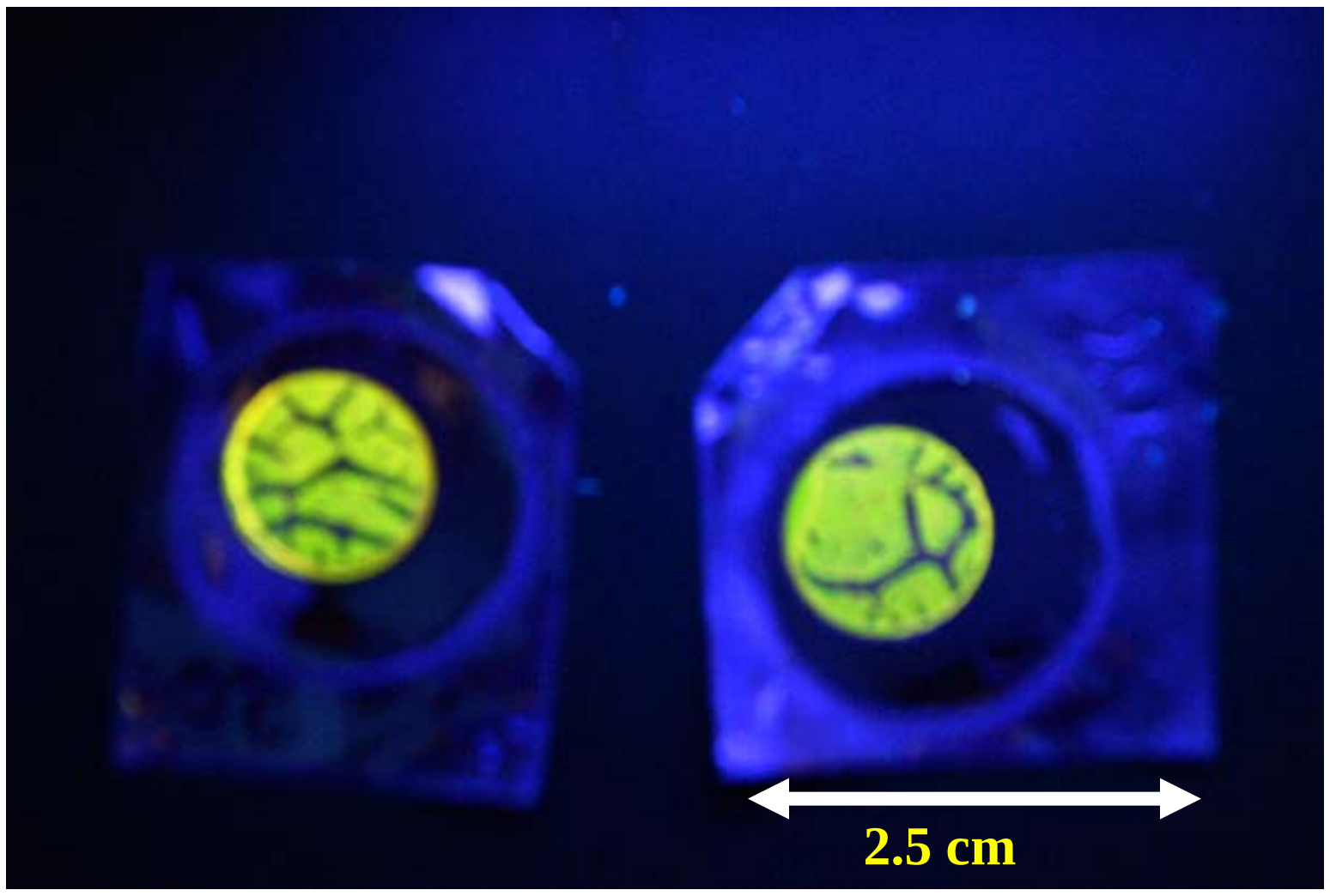

Fig. 3

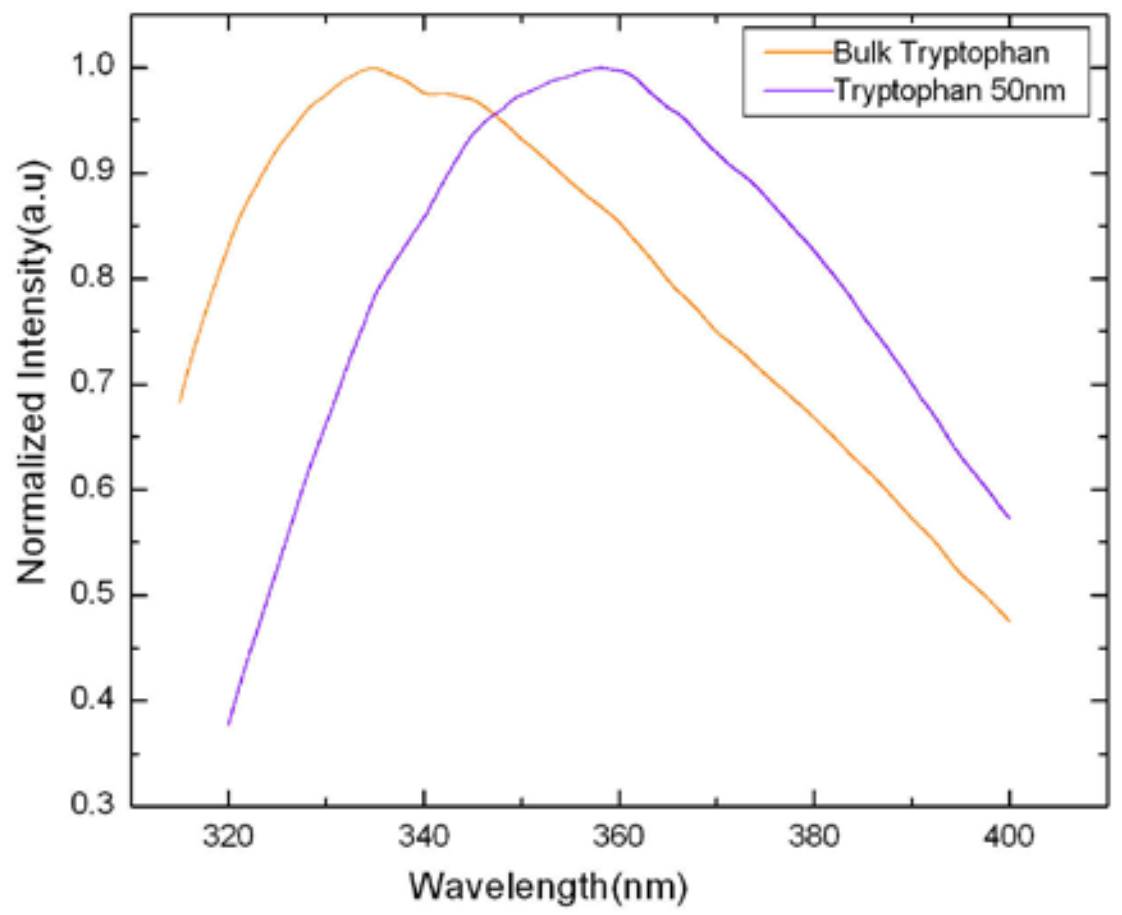

**(a)**

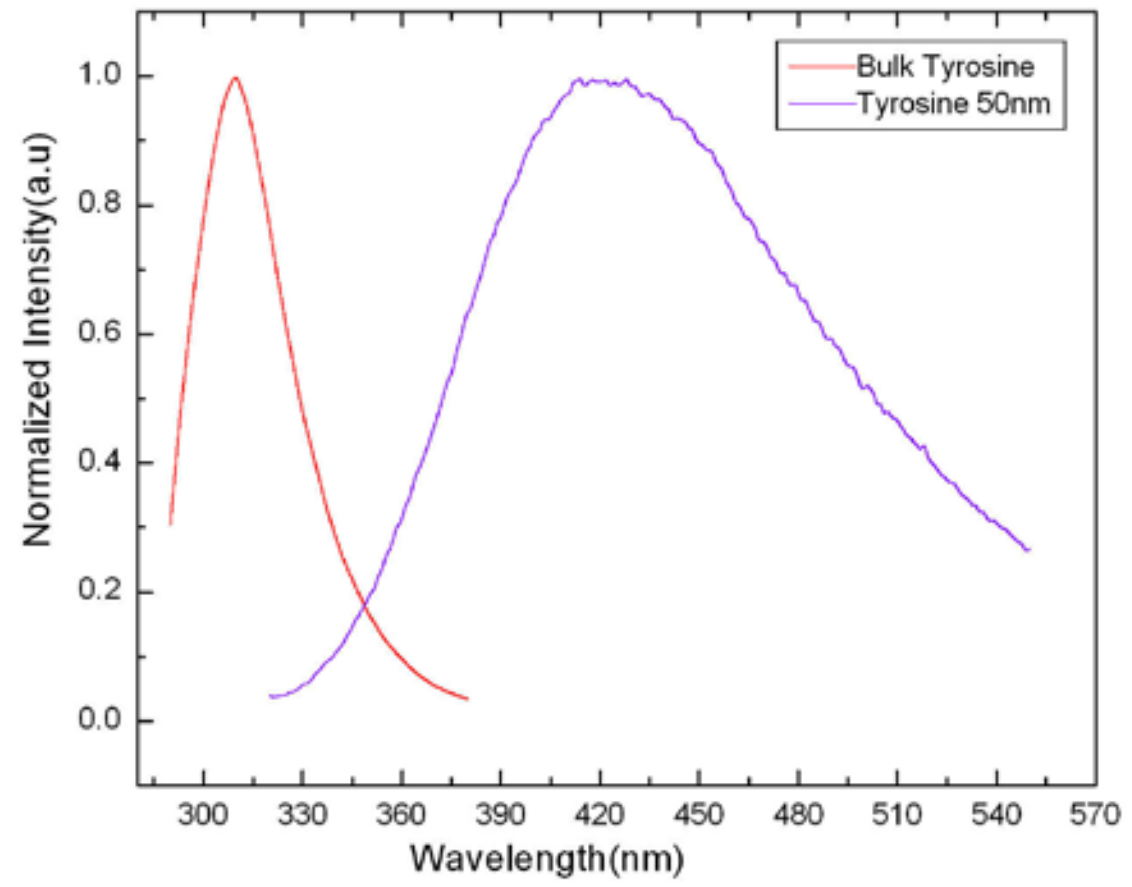

**(b)**

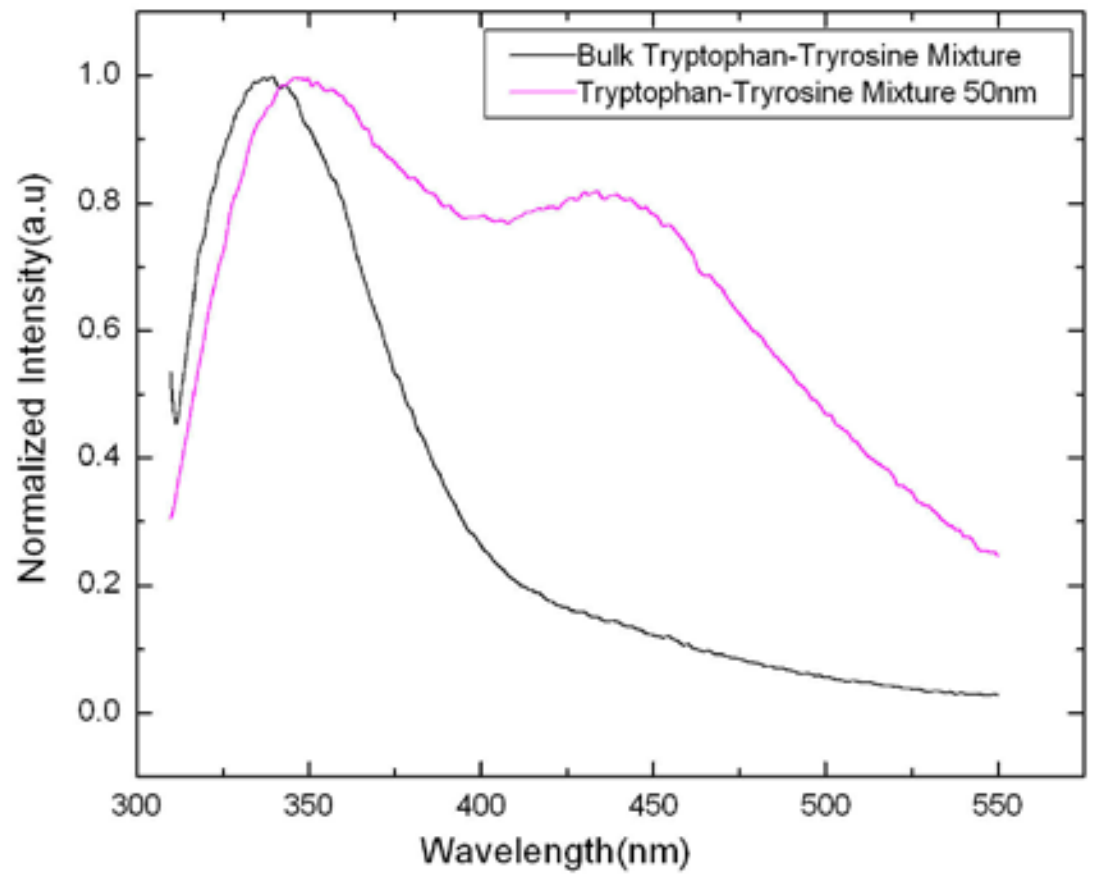

**(c)**

Fig. 4

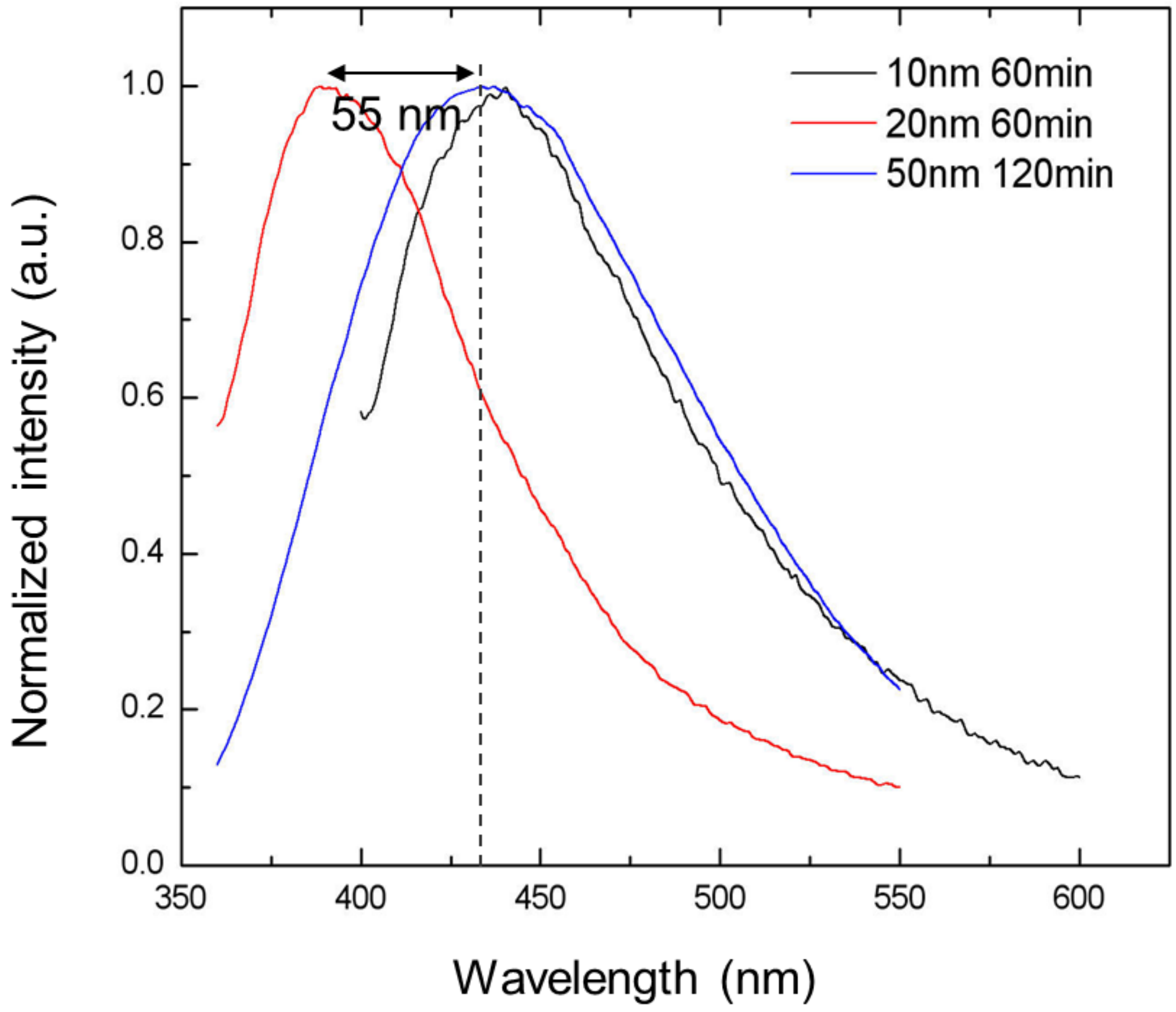

Fig. 5

**Giant increase in the metal-enhanced fluorescence of organic molecules in nanoporous alumina templates and large molecule-specific red/blue shift of the fluorescence peak**

S. Sarkar[1], B. Kanchibotla[2], J. D. Nelson[3], J. D. Edwards[3], J. Anderson[3], G. C. Tepper[1] and S. Bandyopadhyay[2]

[1]Department of Mechanical and Nuclear Engineering, Virginia Commonwealth University, Richmond, Virginia 23284, USA

[2]Department of Electrical and Computer Engineering, Virginia Commonwealth University, Richmond, Virginia 23284, USA

[3]US Army Engineer Research and Development Center, Alexandria, Virginia 22315, USA

# Supporting Information

We provide additional data and micrographs to support to the conclusions in the paper. The authors are indebted to Ms. Jennette Mateo for obtaining Fig. S-3(b).

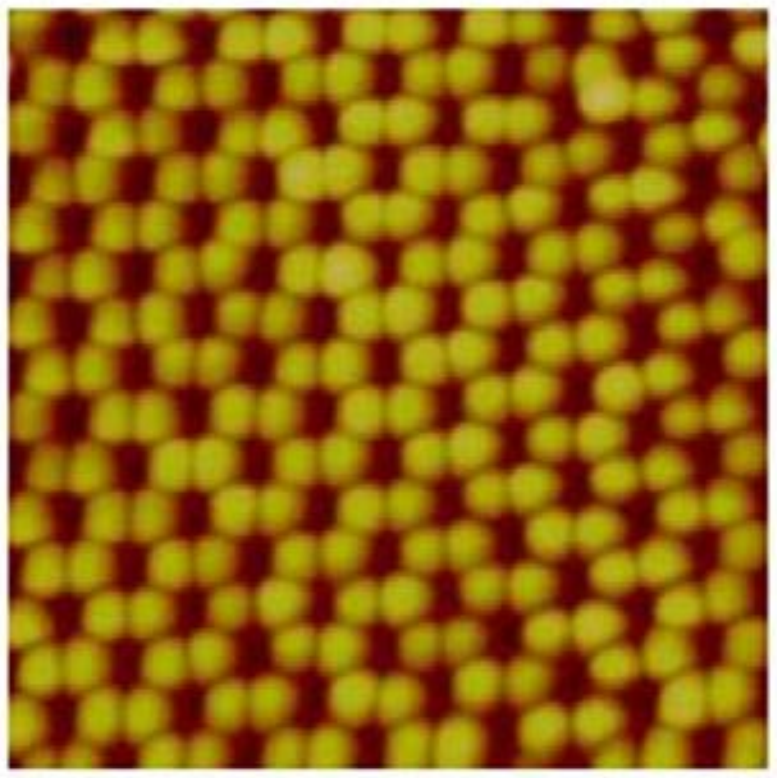

(a)

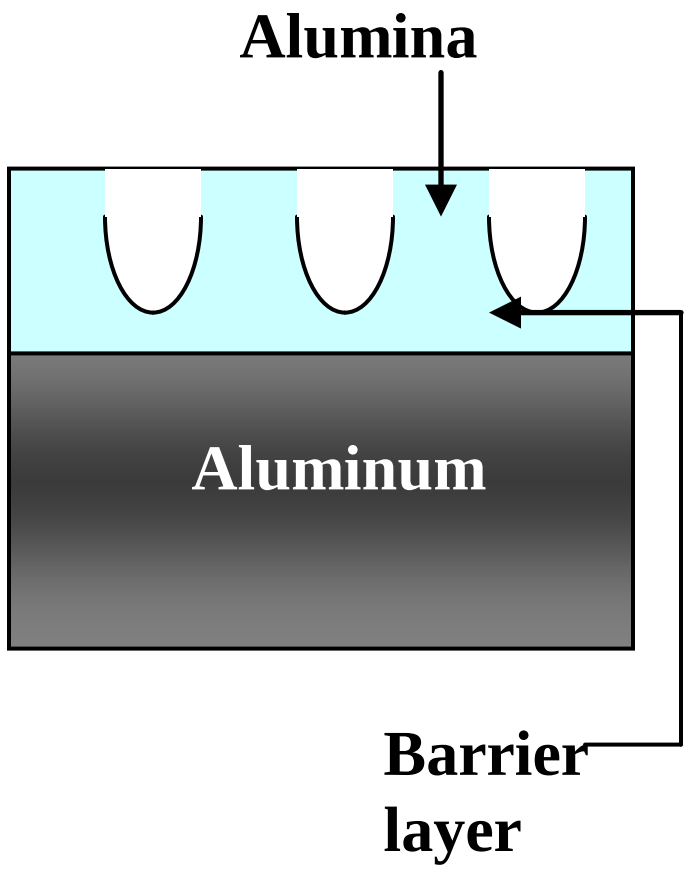

(b)

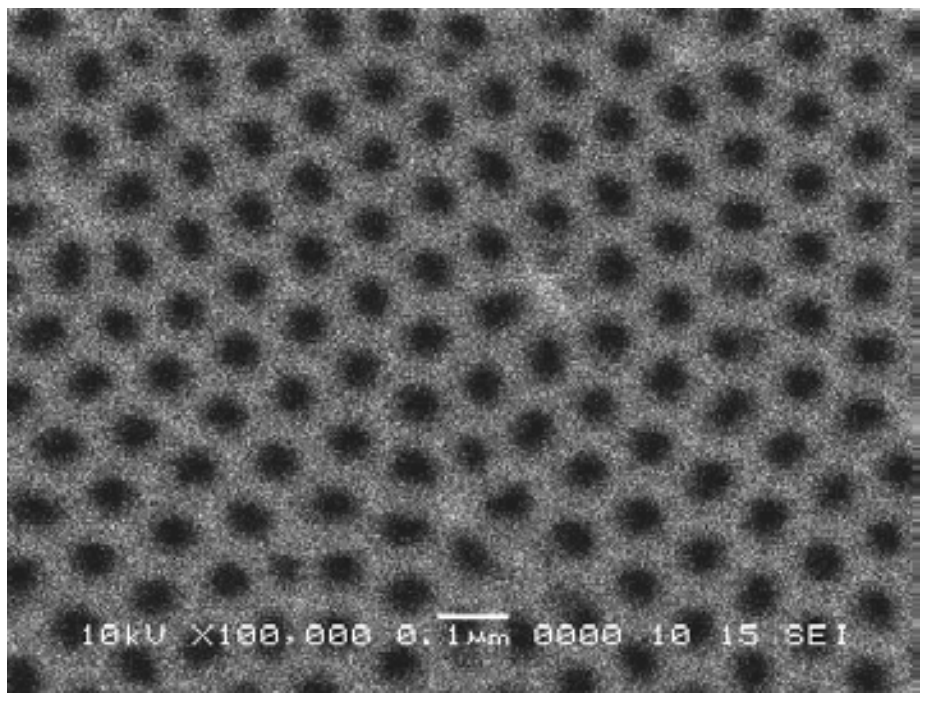

(c)

**Fig. S-1:** An atomic force micrograph of a nanoporous alumina film produced by anodizing Al in 3% oxalic acid. The dark areas are pores and the surrounding light areas are alumina. The pore diameter in this case is 50 nm. (b) Schematic cross sectional view of the structure showing the “barrier layer” (c) To remove the barrier layer, we carried out reverse polarity etching as described in ref. [4]. To ensure that the barrier layer has been completely removed, we peeled off the porous alumina film from the aluminum substrate by dissolving the latter in $HgCl_2$ solution which does not attack alumina. The peeled film was then imaged from the back using SEM. This micrograph shows that the pores have opened up at the bottom indicating successful removal of the barrier layer.

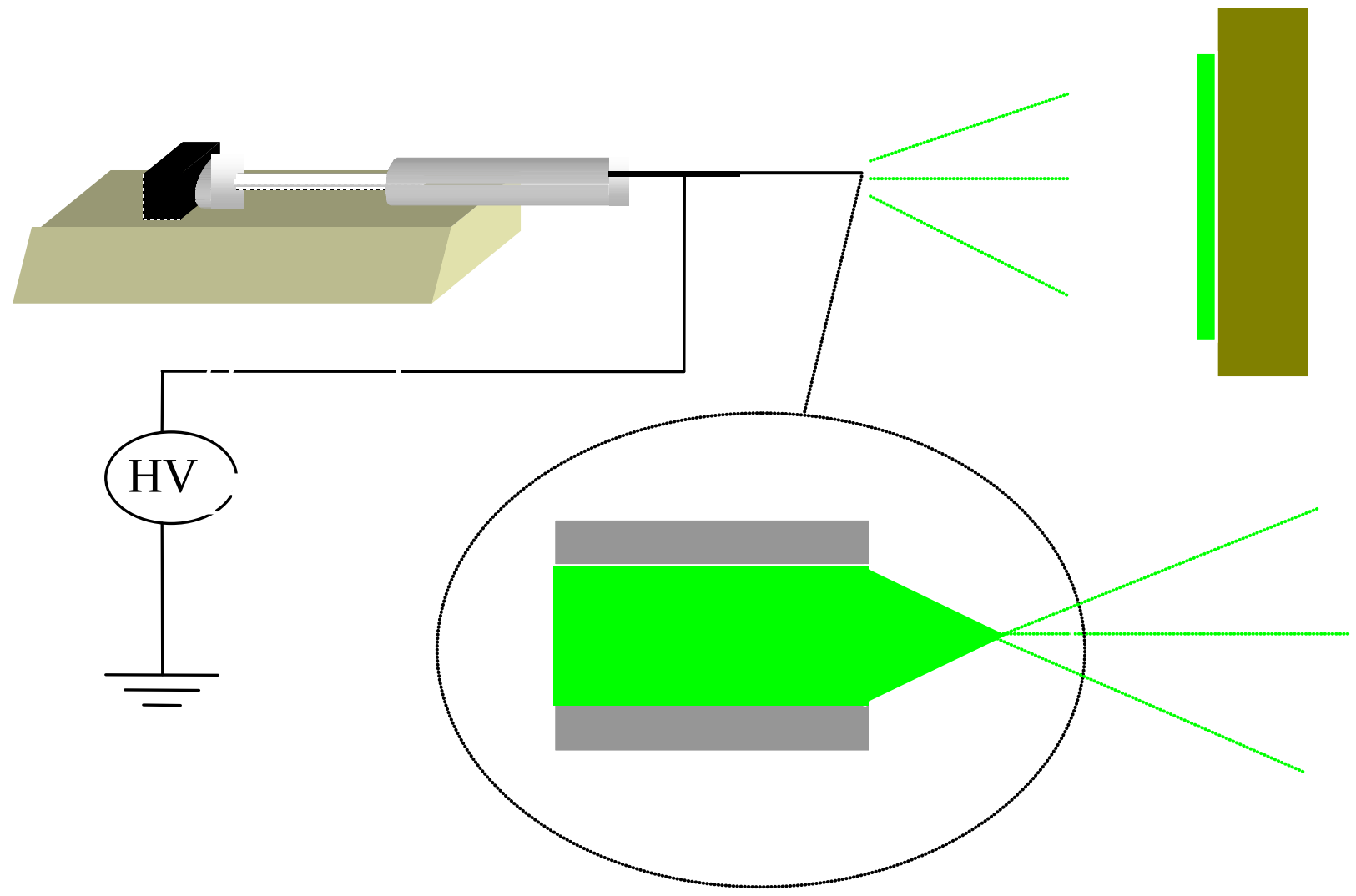

(a)

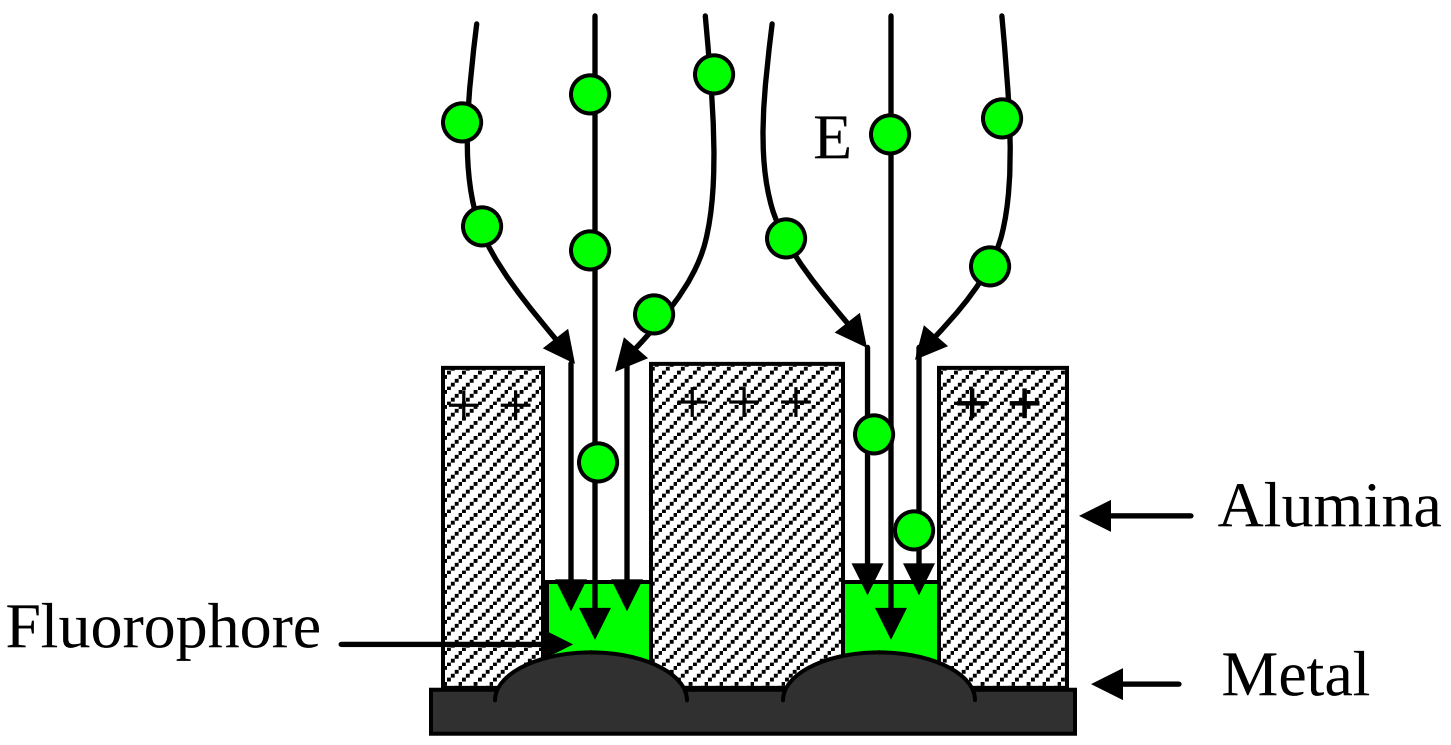

(b)

**Figure S-2:** (a) Schematic diagram of the electospraying process, and (b) illustration of the electric field lines during electrospraying. Note that successful removal of the barrier layer and exposure of the aluminum at the bottom of the pores is essential for *selective* deposition of the molecules within the pores. Since the electric field lines terminate at the bottom of the pores and no electric field line terminates on the surrounding alumina, the fluorophores are deposited only within the pores and not on top of the alumina. The exposed aluminum at the bottom is responsible for the metal enhanced fluorescence that increases the fluorescence intensity.

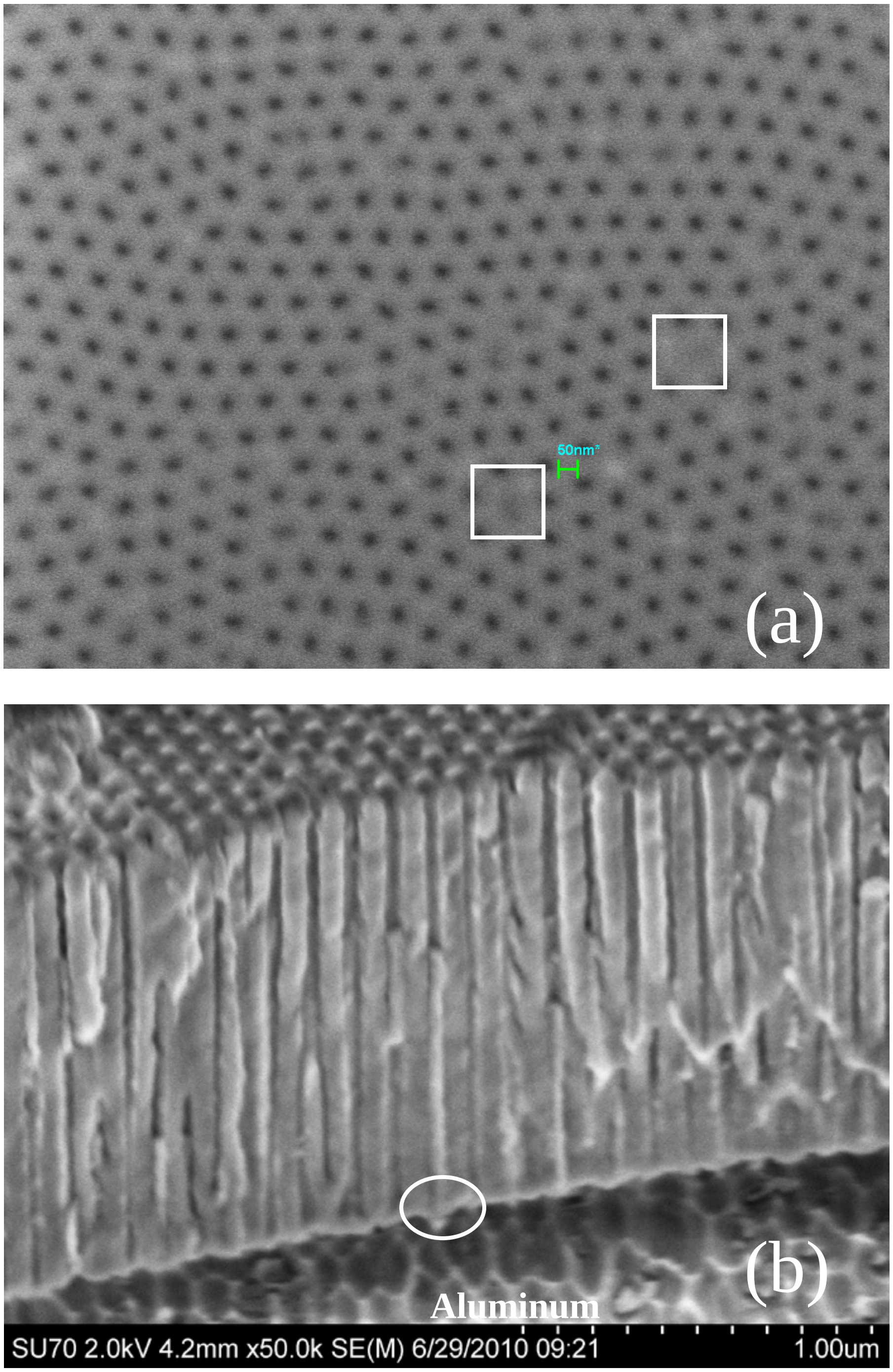

**Fig. S-3**: (a) Scanning electron micrograph of the filled pores showing no overflow. The fluorophores are inside the pores and do not spill over the surface. The pores however are filled somewhat non-uniformly and there are some pores that have been nearly filled to the brim. Two such pores are enclosed within square boxes. (b) Cross-section scanning electron micrograph of the pores filled with a relatively conducting material (CdS) for imaging contrast. The interface between the aluminum substrate and the filler material is indicated with an ellipse.

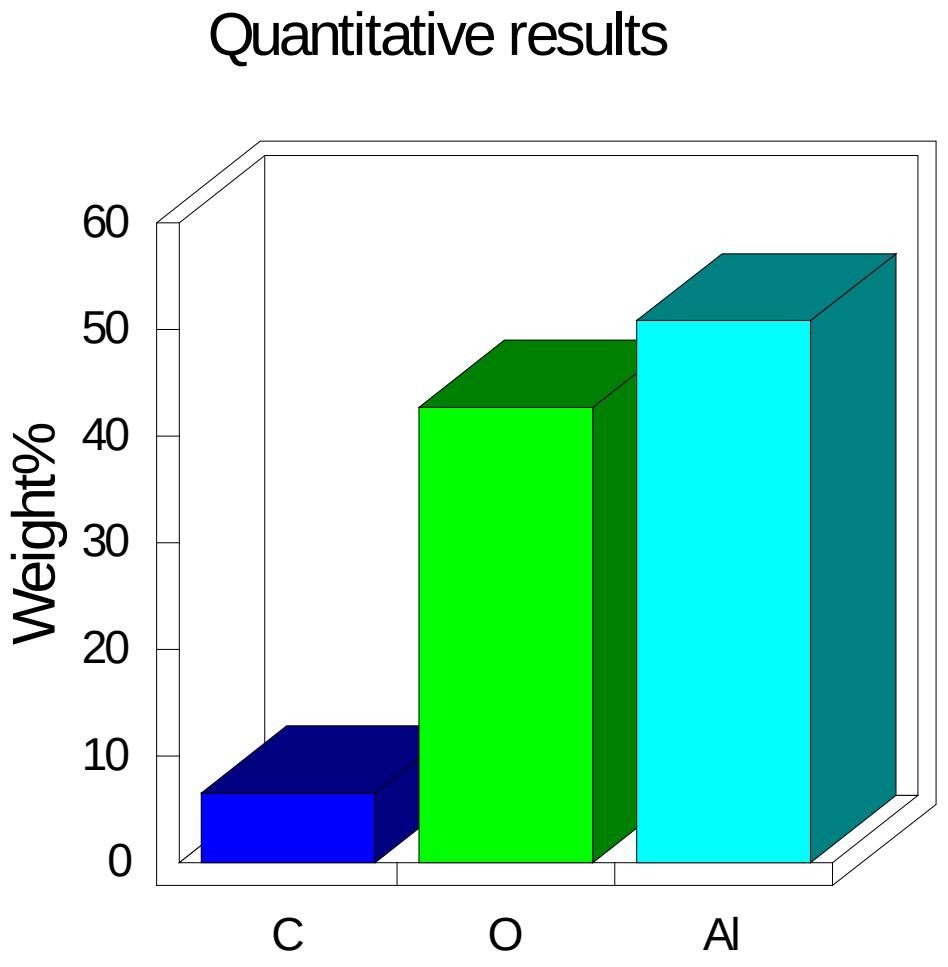

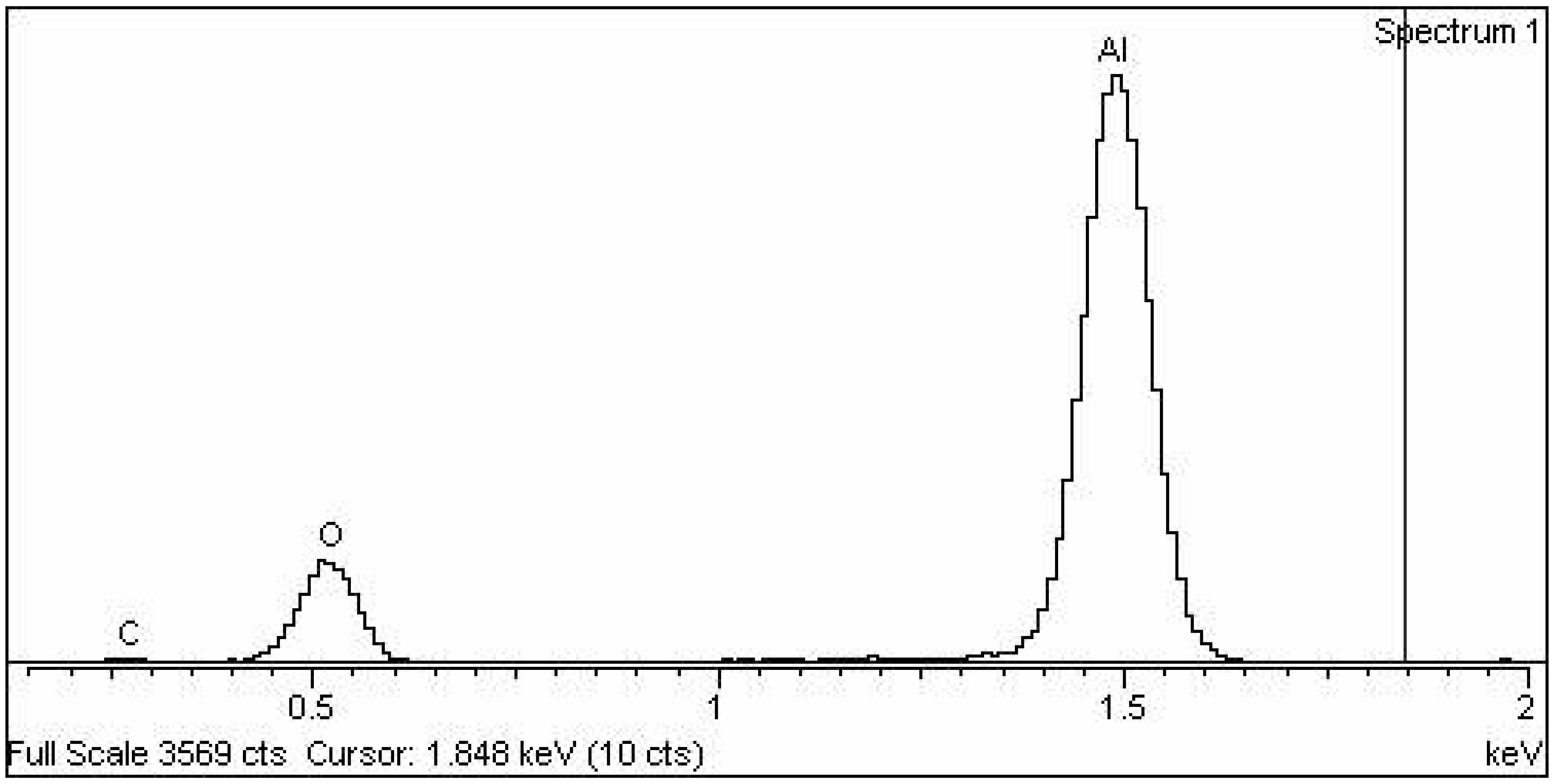

**Fig. S-4:** Energy dispersive analysis of x-ray spectrum. The carbon peak is due to the organic fluorophores resident within the pores. This confirms the presence of fluorophores inside the pores. The Al and O peaks are due to the alumina template.